\documentclass[ aip, jmp,  reprint]{revtex4-1}
\usepackage{amsmath}
\usepackage{graphicx}
\usepackage{dcolumn}
\usepackage{bm}

\begin{document}

\preprint{}
\title[]{Functional relations for the density functional exchange and correlation
functionals connecting functionals at three densities.}
\author{Daniel P. Joubert}
\email{daniel.joubert2@wits.ac.za}
\affiliation{Centre for Theoretical Physics, University of the Witwatersrand, PO Wits
2050, Johannesburg, South Africa}
\date{\today }

\begin{abstract}
It is shown that the DFT exchange and correlation functionals satisfy
\begin{eqnarray}
0 &=&\gamma E_{hx}\left[ \rho _{N}\right] +2E_{c}^{\gamma }\left[ \rho _{N}%
\right]   \notag \\
&&-\gamma E_{hx}\left[ \rho _{N-1}^{\gamma }\right] -2E_{c}^{\gamma }\left[
\rho _{N-1}^{\gamma }\right]   \notag \\
&&+2\int d^{3}r^{\prime }\left( \rho _{N-1}^{0}\left( \mathbf{r}\right)
-\rho _{N-1}^{\gamma }\left( \mathbf{r}\right) \right) v^{0}\left( \left[
\rho _{N}\right] ;\mathbf{r}\right)   \notag \\
&&+\int d^{3}r^{\prime }\left( \rho _{N-1}^{0}\left( \mathbf{r}\right) -\rho
_{N-1}^{\gamma }\left( \mathbf{r}\right) \right) \mathbf{r.\nabla }%
v^{0}\left( \left[ \rho _{N}\right] ;\mathbf{r}\right)   \notag \\
&&+\int d^{3}r^{\prime }\rho _{N}\left( \mathbf{r}\right) \mathbf{r.\nabla }%
 v_{c}^{\gamma }\left( \left[ \rho _{N}\right] ;\mathbf{r}\right) %
   \notag \\
&&-\int d^{3}r^{\prime }\rho _{N-1}^{\gamma }\left( \mathbf{r}\right)
\mathbf{r.\nabla }v_{c}^{\gamma }\left( \left[ \rho _{N-1}^{\gamma }%
\right] ;\mathbf{r}\right)    \notag \\
&&-\int d^{3}r^{\prime }f^{\gamma }\left( \mathbf{r}\right) \mathbf{r.\nabla
}v_{hxc}^{\gamma }\left( \left[ \rho _{N}\right] ;\mathbf{r}\right)   \notag
\\
&&-2\int d^{3}r^{\prime }f^{\gamma }\left( \mathbf{r}\right) v_{hxc}^{\gamma
}\left( \left[ \rho _{N}\right] ;\mathbf{r}\right)   \notag
\end{eqnarray}%
In the derivation of this equation the adiabatic connection formulation is used
where the ground state density of an $N$-electron system, $\rho _{N}$, is kept constant
independent of the electron-electron coupling strength $\gamma$. Here
$E_{hx}\left[ \rho \right]$ is the Hartree plus exchange energy,
$E_{c}^{\gamma }\left[ \rho \right] $ is the correlation energy,
$v_{hxc}^{\gamma }\left[ \rho \right] $ is the Hartree plus exchange-correlation
potential, $v_{c}\left[ \rho \right] $ is the correlation potential and $%
v^{0}\left[ \rho \right] \mathbf{\ }$is the Kohn-Sham potential.  The charge
densities $\rho _{N}$ and $\rho _{N-1}^{\gamma }$ are the $N$- and $\left(
N-1\right) $-electron ground state densities of the same Hamiltonian at
electron-electron coupling strength $\gamma $. $f^{\gamma
}\left( \mathbf{r}\right) =\rho _{N}\left( \mathbf{r}\right) -\rho
_{N-1}^{\gamma }\left( \mathbf{r}\right) $ is the Fukui function. This
equation can be useful in testing the internal self-consistency of
approximations to the exchange and correlation functionals.
\end{abstract}

\maketitle









Density Functional Theory (DFT) \cite{HohenbergKohn:64} is one of the most
important tools for the calculation of electronic structure and structural
properties of molecules and solids. In all practical applications of DFT,
however, approximations to the exact functionals have to be used \cite%
{AdriennRuzsinszky2011,JohnP.Perdew2009,GaborI.Csonka2009a,Staroverov2004,TaoPerdew03}.
Exact relations for density functionals and density functional derivatives
can play an important role in the development of accurate approximations to
the exact functionals. A successful approach to the design of improved
approximate density functionals is by 'constraint satisfaction'\cite%
{JohnP.Perdew2005}, where the approximate functionals are required to
satisfy properties of the exact functionals. The accuracy of approximate
functionals can be tested by comparing to accurate calculations or to
experimental data.  A useful additional test will be to determine the
internal self-consistency of approximations.  With this in mind, the
following expression is derived:%
\begin{eqnarray}
0 &=&\gamma E_{hx}\left[ \rho _{N}\right] +2E_{c}^{\gamma }\left[ \rho _{N}%
\right]   \notag \\
&&-\gamma E_{hx}\left[ \rho _{N-1}^{\gamma }\right] -2E_{c}^{\gamma }\left[
\rho _{N-1}^{\gamma }\right]   \notag \\
&&+2\int d^{3}r^{\prime }\left( \rho _{N-1}^{0}\left( \mathbf{r}\right)
-\rho _{N-1}^{\gamma }\left( \mathbf{r}\right) \right) v^{0}\left( \left[
\rho _{N}\right] ;\mathbf{r}\right)   \notag \\
&&+\int d^{3}r^{\prime }\left( \rho _{N-1}^{0}\left( \mathbf{r}\right) -\rho
_{N-1}^{\gamma }\left( \mathbf{r}\right) \right) \mathbf{r.\nabla }%
v^{0}\left( \left[ \rho _{N}\right] ;\mathbf{r}\right)   \notag \\
&&+\int d^{3}r^{\prime }\rho _{N}\left( \mathbf{r}\right) \mathbf{r.\nabla }%
 v_{c}^{\gamma }\left( \left[ \rho _{N}\right] ;\mathbf{r}\right) %
   \notag \\
&&-\int d^{3}r^{\prime }\rho _{N-1}^{\gamma }\left( \mathbf{r}\right)
\mathbf{r.\nabla } v_{c}^{\gamma }\left( \left[ \rho _{N-1}^{\gamma }%
\right] ;\mathbf{r}\right)    \notag \\
&&-\int d^{3}r^{\prime }f^{\gamma }\left( \mathbf{r}\right) \mathbf{r.\nabla
}v_{hxc}^{\gamma }\left( \left[ \rho _{N}\right] ;\mathbf{r}\right)   \notag
\\
&&-2\int d^{3}r^{\prime }f^{\gamma }\left( \mathbf{r}\right) v_{hxc}^{\gamma
}\left( \left[ \rho _{N}\right] ;\mathbf{r}\right)   \label{res0}
\end{eqnarray}%
In the derivation of this expression the adiabatic connection formulation\cite%
{HarrisJones:74,LangrethPerdew:75,LangrethPerdew:77,GunnarsonLundqvist:76}
is used where the ground state density of an $N$-electron system, $\rho _{N}$,
is kept constant independent of the electron-electron coupling strength $%
\gamma$. In Eq.(\ref{res0}) $v_{hxc}^{\gamma }\left[ \rho \right] $ is the Hartree
plus exchange-correlation potential, $v_{c}\left[ \rho \right] $ is the
correlation potential and $v^{0}\left[ \rho \right] \mathbf{\ }$is the
Kohn-Sham potential. $E_{c}^{\gamma }\left[ \rho \right] $ is the
correlation energy while $E_{hx}\left[ \rho \right] $ is the Hartree plus
exchange energy. The charge densities $\rho _{N}$ and $\rho _{N-1}^{\gamma }$
are the $N$- and $\left( N-1\right) $-electron ground state densities of the
same Hamiltonian at electron-electron coupling strength $\gamma $ as
discussed below. $f^{\gamma }\left( \mathbf{r}\right) =\rho _{N}\left(
\mathbf{r}\right) -\rho _{N-1}^{\gamma }\left( \mathbf{r}\right) $ is the
Fukui function.

Equation (\ref{res0}) couples functionals and functional derivatives
evaluated at three different densities and two particle numbers. This
equation can be used as a stringent test to check the internal
self-consistency of approximations to the exchange and correlation
functionals. If exact exchange is used it can serve as a check on
approximations to the correlation energy functional.

Evaluation of (\ref{res0}) requires two independent self-consistent
Kohn-Sham calculations, one for the $N$-electron system to
determine $\rho _{N}$ and $\rho _{N-1}^{0},$ and one for the $\left(
N-1\right) $-electron system to determine $\rho _{N-1}^{\gamma }.$ Once the
densities are found, the functionals and functional derivatives can be used
in Eq.(\ref{res0}) as a test of how well the approximations work.

\section{Proof of Equation (\ref{res0})}

According to the Hohenberg-Kohn theorem \cite{HohenbergKohn:64} and its
generalization to degenerate states\cite{Levy:79} the ground state energy $%
E^{\gamma }\left[ \rho \right] $ of a system of interacting electrons is a
functional of $\rho $, the ground state density \cite%
{ParrYang:bk89,DreizlerGross:bk90}.%
\begin{equation}
E^{\gamma }\left[ \rho \right] =T^{\gamma }\left[ \rho \right] +\gamma
V_{ee}^{\gamma }\left[ \rho \right] +\int d^{3}r\rho \left( \mathbf{r}%
\right) v^{\gamma }\left( \left[ \rho \right] ;\mathbf{r}\right)
\label{ks0}
\end{equation}%
where $T^{\gamma }\left[ \rho \right] $ is the kinetic energy and $\gamma
V_{ee}^{\gamma }\left[ \rho \right] $ the mutual Coulomb interaction energy
at density $\rho .$ In order to derive Eq.(\ref {res0}) the adiabatic connection
approach will be used \cite%
{HarrisJones:74,LangrethPerdew:75,LangrethPerdew:77,GunnarsonLundqvist:76}
in which the external potential $v_{\text{ext}}^{\gamma }\left( \left[ \rho %
\right] ;\mathbf{r}\right) $ is constructed to keep the ground state density
independent of the electron-electron interaction strength, scaled by $\gamma
,$ and has the form \cite{LevyPerdew:85,GorlingLevy:93}
\begin{align}
v^{\gamma }(\left[ \rho \right] ;\mathbf{r})& =\left( 1-\gamma \right)
v_{hx}([\rho ];\mathbf{r})  \notag \\
& +v_{c}^{1}([\rho ];\mathbf{r)}-v_{c}^{\gamma }([\rho ];\mathbf{r)+}v_{%
\text{ext}}(\mathbf{r}).  \label{e1}
\end{align}%
$v^{1}(\left[ \rho \right] ;\mathbf{r})=v_{\text{ext}}\left( \mathbf{r}%
\right) $ is the external potential at full coupling strength, $\gamma =1,$
and $v^{0}(\left[ \rho \right] ;\mathbf{r})$ is non-interacting Kohn-Sham
potential. The exchange plus Hartree potential \cite%
{ParrYang:bk89,DreizlerGross:bk90} $v_{hx}([\rho ];\mathbf{r})$, is
independent of $\gamma ,$ while the correlation potential $v_{c}^{\gamma
}([\rho ];\mathbf{r)}$ depends in the scaling parameter $\gamma.$ The
adiabatically scaled $N$-electron Hamiltonian $\hat{H}^{\gamma }$ has the
form \cite%
{HarrisJones:74,LangrethPerdew:75,LangrethPerdew:77,GunnarsonLundqvist:76}
\begin{equation}
\hat{H}^{\gamma }=\hat{T}+\gamma \hat{V}_{ee}+\hat{v}^{\gamma }\left[ \rho %
\right] .  \label{a3}
\end{equation}%
Atomic units, $\hbar =e=m=1$ are used throughout. $\hat{T}$ is the kinetic
energy operator,%
\begin{equation}
\hat{T}=-\frac{1}{2}\sum_{i=1}^{N}\nabla _{i}^{2},  \label{a4}
\end{equation}%
\ and $\gamma \hat{V}_{\text{ee }}$is a scaled electron-electron interaction,%
\begin{equation}
\gamma \hat{V}_{ee}=\gamma \sum_{i<j}^{N}\frac{1}{\left\vert \mathbf{r}_{i}-%
\mathbf{r}_{j}\right\vert }.  \label{a2}
\end{equation}%
and
\begin{equation}
\hat{v}^{\gamma }\left[ \rho \right] =\sum_{i=1}^{N}v^{\gamma }\left[ \rho %
\right] ;\mathbf{r}_{i}.  \label{a0}
\end{equation}

Define the energy functional\cite{ParrYang:bk89,DreizlerGross:bk90}%
\begin{eqnarray}
F^{\gamma }\left[ \rho \right] &=&T^{\gamma }\left[ \rho \right] +\gamma
V_{ee}^{\gamma }\left[ \rho \right]  \notag \\
&=&T^{0}\left[ \rho \right] +\gamma E_{hx}\left[ \rho \right] +E_{c}^{\gamma
}\left[ \rho \right]  \label{F}
\end{eqnarray}%
where $E_{hx}\left[ \rho \right] $ is the Hartree plus exchange
energy, and $E_{c}^{\gamma }\left[ \rho \right] $ is the correlation energy.
Note that $E_{hx}\left[ \rho \right] $ is independent of $\gamma$. As shown
in appendix \ref{appendixA}%
\begin{eqnarray}
\mu  &=&v^{\gamma }\left( \mathbf{r}\right) +\frac{1}{2}\mathbf{r.\nabla }%
v^{\gamma }\left( \mathbf{r}\right) +\frac{1}{2}\left( v_{hxc}^{\gamma
}\left( \mathbf{r}\right) -t_{c}^{\gamma }\left( \mathbf{r}\right) \right)
\notag \\
&&-\frac{1}{2}\int d^{3}r^{\prime }\rho \left( \mathbf{r}^{\prime }\right)
\mathbf{r}^{\prime }\mathbf{.\nabla }^{\prime }\frac{\delta ^{2}F^{\gamma }%
\left[ \rho \right] }{\delta \rho \left( \mathbf{r}^{\prime }\right) \delta
\rho \left( \mathbf{r}\right) },  \label{ip4}
\end{eqnarray}%
where $t_{c}^{\gamma }\left( \left[ \rho \right] ;\mathbf{r}\right) =\frac{%
\delta T_{c}^{\gamma }\left[ \rho \right] }{\delta \rho \left( \mathbf{r}%
\right) }$ with $T_{c}^{\gamma }\left[ \rho \right] $ the correlation part
of the kinetic energy and $v_{hxc}^{\gamma }\left( \mathbf{r}\right) =\gamma
v_{hx}\left( \mathbf{r}\right) +v_{c}^{\gamma }\left( \mathbf{r}\right) $.
For notational convenience the functional dependence on $\rho $ has been
suppressed. The chemical potential $\mu $ depends on the asymptotic decay of
the charge density\cite{LevyGorling:96,LevyGorlingb:96}, and hence, since
the charge density $\rho $ is independent of $\gamma $ by construction, $\mu
$ is independent of $\gamma $.

The Fukui function \cite{MaxBerkowitz1988,P.Fuentalba2007}%
\begin{equation}
f^{\gamma }\left( \mathbf{r}\right) =\left. \frac{\delta \mu }{\delta
v^{\gamma }\left( \mathbf{r}\right) }\right\vert _{N}=\left. \frac{\delta
\rho \left( \mathbf{r}\right) }{\delta N}\right\vert _{v^{\gamma }}
\label{bp6}
\end{equation}%
satisfies%
\begin{equation}
\int d^{3}r^{\prime }\frac{\delta ^{2}F^{\gamma }\left[ \rho \right] }{%
\delta \rho \left( \mathbf{r}\right) \delta \rho \left( \mathbf{r}^{\prime
}\right) }f^{\gamma }\left( \mathbf{r}^{\prime }\right) =\left. \frac{%
\partial \mu }{\partial N}\right\vert _{v^{\gamma }}  \label{bp7}
\end{equation}%
and
\begin{equation}
\int d^{3}r^{\prime }f^{\gamma }\left( \mathbf{r}^{\prime }\right) =1.
\label{bp8}
\end{equation}%
Since%
\begin{equation}
\int d^{3}r^{\prime }\left( 3\rho \left( \mathbf{r}^{\prime }\right) +%
\mathbf{r}^{\prime \prime }\mathbf{.\nabla }\rho \left( \mathbf{r}^{\prime
}\right) \right) =0,  \label{d0}
\end{equation}%
it follows from (\ref{ip4}), (\ref{bp7}) and (\ref{bp8}) that
\begin{eqnarray}
\mu  &=&\int d^{3}r^{\prime }f^{\gamma }\left( \mathbf{r}\right) v^{\gamma
}\left( \mathbf{r}\right) +\frac{1}{2}\int d^{3}r^{\prime }f^{\gamma }\left(
\mathbf{r}\right) \mathbf{r.\nabla }v^{\gamma }\left( \mathbf{r}\right) +
\notag \\
&&\frac{1}{2}\int d^{3}r^{\prime }f^{\gamma }\left( \mathbf{r}\right) \left(
v_{hxc}^{\gamma }\left( \mathbf{r}\right) -t_{c}^{\gamma }\left( \mathbf{r}%
\right) \right) .  \label{d1}
\end{eqnarray}%
Using (\ref{e1}), this equation can be reduced to%
\begin{eqnarray}
&&\int d^{3}r^{\prime }f^{\gamma }\left( \mathbf{r}\right) \left[
2v^{0}\left( \mathbf{r}\right) +\mathbf{r.\nabla }v^{0}\left( \mathbf{r}%
\right) \right] -2\mu   \notag \\
&=&\int d^{3}r^{\prime }f^{\gamma }\left( \mathbf{r}\right) \mathbf{r.\nabla
}v_{hxc}^{\gamma }\left( \mathbf{r}\right)   \notag \\
&&+\int d^{3}r^{\prime }f^{\gamma }\left( \mathbf{r}\right) \left[
v_{hxc}^{\gamma }\left( \mathbf{r}\right) +t_{c}^{\gamma }\left( \mathbf{r}%
\right) \right].  \label{d3}
\end{eqnarray}%
From the virial theorem for the Kohn-Sham independent particle wavefunctions%
\cite{Merzbacher1970}%
\begin{eqnarray}
&&\int d^{3}r^{\prime }\left[ \rho _{N}\left( \mathbf{r}\right) -\rho
_{N-1}^{0}\left( \mathbf{r}\right) \right] \left[ 2v^{0}\left( \mathbf{r}%
\right) +\mathbf{r.\nabla }v^{0}\left( \mathbf{r}\right) \right]   \notag \\
&=&2\left( E^{0}\left[ \rho \right] -E^{0}\left[ \rho _{N-1}^{0}\right]
\right) ,  \label{d4}
\end{eqnarray}%
where $E^{0}\left[ \rho \right] $ and $E^{0}\left[ \rho _{N-1}^{0}\right] $
are the Kohn-Sham ground state energies of the independent $N$- and $\left(
N-1\right) $-particle systems of the same Kohn-Sham Hamiltonian with potential $v^{0}$%
. By definition, since $\mu $ is independent of $\gamma $\cite%
{LevyGorling:96,LevyGorlingb:96}$,$
\begin{equation}
\mu =E^{\gamma }\left[ \rho \right] -E^{\gamma }\left[ \rho _{N-1}^{\gamma }%
\right]   \label{mu}
\end{equation}%
where $E^{\gamma }\left[ \rho _{N}\right] $ and $E^{\gamma }\left[ \rho
_{N-1}^{\gamma }\right] $ are the ground state energies of the $N$- and $%
\left( N-1\right) $-particle systems of the same Hamiltonian $H^{\gamma }$,
Eq. (\ref{a3}), with $\rho _{N}$ and $\rho _{N-1}^{\gamma }$ the
corresponding ground state densities. By construction $\rho _{N}^{\gamma } =\rho _{N}$ is
independent of $\gamma $ but $\rho _{N-1}^{\gamma }$ is expected to be a
function of $\gamma .$

Since the Fukui function \cite{PPLB:82,P.Geerlings2003}
\begin{eqnarray}
f^{\gamma }\left( \mathbf{r}\right)  &=&\left. \frac{\delta \rho \left(
\mathbf{r}\right) }{\delta N}\right\vert _{v^{\gamma }}  \notag \\
&=&\rho _{N}\left( \mathbf{r}\right) -\rho _{N-1}^{\gamma }\left( \mathbf{r}%
\right)   \label{f2}
\end{eqnarray}%
is the difference between the ground state densities of the $N$- particle
and $\left( N-1\right) $-particle systems, it follows from (\ref{d3}), (\ref%
{d4}) and (\ref{mu}), that%
\begin{eqnarray}
&&2\int d^{3}r^{\prime }\left( \rho _{N-1}^{0}\left( \mathbf{r}\right) -\rho
_{N-1}^{\gamma }\left( \mathbf{r}\right) \right) v^{0}\left( \mathbf{r}%
\right) +  \notag \\
&&+\int d^{3}r^{\prime }\left( \rho _{N-1}^{0}\left( \mathbf{r}\right) -\rho
_{N-1}^{\gamma }\left( \mathbf{r}\right) \right) \mathbf{r.\nabla }%
v^{0}\left( \mathbf{r}\right)   \notag \\
&=&\int d^{3}r^{\prime }f^{\gamma }\left( \mathbf{r}\right) \mathbf{r.\nabla
}v_{hxc}^{\gamma }\left( \mathbf{r}\right)   \notag \\
&&+\int d^{3}r^{\prime }f^{\gamma }\left( \mathbf{r}\right) \left[
v_{hxc}^{\gamma }\left( \mathbf{r}\right) +t_{c}^{\gamma }\left( \mathbf{r}%
\right) \right]   \label{d5}
\end{eqnarray}%
The reference to the correlation part of the kinetic energy can be
eliminated as follows. Since\cite{LevyPerdew:85}
\begin{equation}
\gamma V_{ee}^{\gamma }\left[ \rho \right] =\gamma E_{hx}\left[ \rho \right]
+E_{c}^{\gamma }\left[ \rho \right] -T_{c}^{\gamma }\left[ \rho \right] ,
\label{d5a}
\end{equation}%
the last line in (\ref{d5}) can be written as%
\begin{eqnarray}
&&\int d^{3}r^{\prime }f^{\gamma }\left( \mathbf{r}\right) \left[
v_{hxc}^{\gamma }\left( \mathbf{r}\right) +t_{c}^{\gamma }\left( \mathbf{r}%
\right) \right]   \notag \\
&=&-\gamma \int d^{3}r^{\prime }f^{\gamma }\left( \mathbf{r}\right) \frac{%
\delta V_{ee}^{\gamma }\left[ \rho \right] }{\delta \rho \left( \mathbf{r}%
\right) }  \notag \\
&&+2\int d^{3}r^{\prime }f^{\gamma }\left( \mathbf{r}\right) v_{hxc}^{\gamma
}\left( \mathbf{r}\right)   \label{d6}
\end{eqnarray}%
As shown in appendix \ref{appendixB}\cite{Joubert2011a},%
\begin{equation}
\int d^{3}r^{\prime }f^{\gamma }\left( \mathbf{r}\right) \frac{\delta
V_{ee}^{\gamma }\left[ \rho _{N}\right] }{\delta \rho _{N}\left( \mathbf{r}%
\right) }=V_{ee}^{\gamma }\left[ \rho _{N}\right] -V_{ee}^{\gamma }\left[
\rho _{N-1}^{\gamma }\right] .  \label{Vee}
\end{equation}%
Since \cite{LevyPerdew:85}%
\begin{equation}
E_{c}^{\gamma }\left[ \rho \right] +T_{c}^{\gamma }\left[ \rho \right]
=-\int d^{3}r^{\prime }\rho \left( \mathbf{r}\right) \mathbf{r.\nabla }\left[
v_{c}^{\gamma }\left( \left[ \rho \right] ;\mathbf{r}\right) \right] ,
\label{vtc}
\end{equation}%
it follows from (\ref{d5a}), (\ref{d6}), (\ref{Vee}) and (\ref{vtc}) that%
\begin{eqnarray}
&&\int d^{3}r^{\prime }f^{\gamma }\left( \mathbf{r}\right) \left[
v_{hxc}^{\gamma }\left( \mathbf{r}\right) +t_{c}^{\gamma }\left( \mathbf{r}%
\right) \right]   \notag \\
&&+\int d^{3}r^{\prime }\rho _{N}\left( \mathbf{r}\right) \mathbf{r.\nabla }%
v_{c}^{\gamma }\left( \left[ \rho _{N}\right] ;\mathbf{r}\right)   \notag \\
&&-\int d^{3}r^{\prime }\rho _{N-1}^{\gamma }\left( \mathbf{r}\right)
\mathbf{r.\nabla }v_{c}^{\gamma }\left( \left[ \rho _{N-1}^{\gamma }\right] ;%
\mathbf{r}\right)   \notag \\
&=&\gamma E_{hx}\left[ \rho _{N-1}^{\gamma }\right] +2E_{c}^{\gamma }\left[
\rho _{N-1}^{\gamma }\right]   \notag \\
&&-\gamma E_{hx}\left[ \rho _{N}\right] -2E_{c}^{\gamma }\left[ \rho _{N}%
\right]   \notag \\
&&+2\int d^{3}r^{\prime }f^{\gamma }\left( \mathbf{r}\right) v_{hxc}^{\gamma
}\left( \mathbf{r}\right) .  \label{d7}
\end{eqnarray}%
Combining (\ref{d5}) and (\ref{d7}) leads to Equation (\ref{res0}).

This equation is now entirely in terms of the exchange and correlation
energy and potentials and other known density dependent quantities.

\section{Discussion and Summary}

Equation(\ref{res0}) is valid for pure state DFT \cite%
{ParrYang:bk89,DreizlerGross:bk90} since equations (\ref{d4}) and (\ref{Vee}%
) have been shown to be correct for pure states.  Throughout it was assumed
that the functionals derivatives are well defined \cite%
{ParrYang:bk89,DreizlerGross:bk90,Lieb1983}.

Note that  $\rho _{N-1}^{\gamma }\neq \rho _{N-1}^{0}.$ In order to
determine $\rho _{N-1}^{\gamma }$ a Kohn-Sham calculation with\cite%
{GorlingLevyB:95}
\begin{equation*}
v^{0}\left[ \rho _{N-1}^{\gamma }\right] =v^{\gamma }\left[ \rho _{N}\right]
+\gamma v_{hx}\left[ \rho _{N-1}^{\gamma }\right] +v_{c}^{\gamma }\left[
\rho _{N-1}^{\gamma }\right]
\end{equation*}%
has to be performed. Note that $v^{\gamma }\left[ \rho _{N-1}^{\gamma }%
\right] =v^{\gamma }\left[ \rho _{N}\right] .$

In summary, an equation that couples exchange and correlation functionals
and functional derivatives evaluated at three different densities and for two
particle numbers has been derived. This equation can be used as a stringent
test to check the internal self-consistency of approximations to the
exchange and correlation functionals.

\section*{Acknowledgements}

The author acknowledges support from the National Research Foundation (NRF).

\bigskip \appendix

\section{\label{appendixA}Derivation of Eq. (\protect\ref{ip4})}

Let $\rho _{\lambda }\left( \mathbf{r}\right) =\lambda ^{3}\rho \left(
\lambda \mathbf{r}\right) ,$ the uniformly scaled density. Then (see for
example Eq.(A.33) in Appendix A of reference \cite{ParrYang:bk89}),

\begin{eqnarray}
&&\frac{d}{d\lambda }\left. \frac{\delta F^{^{\gamma }}\left[ \rho _{\lambda
}\right] }{\delta \rho _{\lambda }\left( \mathbf{r}\right) }\right\vert
_{\rho _{\lambda },\lambda =1}  \notag \\
&=&\int d^{3}r^{\prime }\left( 3\rho \left( \mathbf{r}^{\prime }\right) +%
\mathbf{r}^{\prime }\mathbf{.\nabla }^{\prime }\rho \left( \mathbf{r}%
^{\prime }\right) \right) \frac{\delta ^{2}F^{^{\gamma }}\left[ \rho \right]
}{\delta \rho \left( \mathbf{r}^{\prime }\right) \delta \rho \left( \mathbf{r%
}\right) }  \notag \\
&=&-\int d^{3}r^{\prime }\rho \left( \mathbf{r}^{\prime }\right) \mathbf{r}%
^{\prime }\mathbf{.\nabla }^{\prime }\frac{\delta ^{2}F^{^{\gamma }}\left[
\rho \right] }{\delta \rho \left( \mathbf{r}^{\prime }\right) \delta \rho
\left( \mathbf{r}\right) }.  \label{ip2}
\end{eqnarray}%
The last line is valid if $\rho \left( \mathbf{r}^{\prime }\right) $
vanishes when $r\rightarrow \infty $ as would be the case for a finite
system. Now consider the Schr\"{o}dinger equation%
\begin{eqnarray}
&&\left[ -\frac{1}{2}\sum_{i=1}^{N}\nabla _{i}^{2}+\frac{\gamma }{\lambda }%
\sum_{i<j}^{N}\frac{1}{\left\vert \mathbf{r}_{i}-\mathbf{r}_{j}\right\vert }%
\right. +  \notag \\
&&\left. \sum_{i=1}^{N}v^{\frac{^{\gamma }}{\lambda }}\left( \left[ \rho %
\right] ;\mathbf{r}_{i}\right) \right] \Psi ^{\frac{^{\gamma }}{\lambda }%
}\left( \left\{ \mathbf{r}_{i}\right\} \right)   \notag \\
&=&E^{\frac{\gamma }{\lambda }}\left( v^{\frac{^{\gamma }}{\lambda }}\left[
\rho \right] \right) \Psi ^{\frac{^{\gamma }}{\lambda }}\left( \left\{
\mathbf{r}_{i}\right\} \right)   \label{s11}
\end{eqnarray}%
from which it follows that%
\begin{eqnarray}
&&\left[ -\frac{1}{2}\sum_{i=1}^{N}\nabla _{i}^{2}+\gamma \sum_{i<j}^{N}%
\frac{1}{\left\vert \mathbf{r}_{i}-\mathbf{r}_{j}\right\vert }+\right.
\notag \\
&&\left. \sum_{i=1}^{N}\lambda ^{2}v^{\frac{^{\gamma }}{\lambda }}\left( %
\left[ \rho \right] ;\mathbf{r}_{i}\right) \right] \Psi ^{\frac{^{\gamma }}{%
\lambda }}\left( \left\{ \lambda \mathbf{r}_{i}\right\} \right)   \notag \\
&=&\lambda ^{2}E^{\frac{\gamma }{\lambda }}\left( v^{\frac{^{\gamma }}{%
\lambda }}\left[ \rho \right] \right) \Psi ^{\frac{^{\gamma }}{\lambda }%
}\left( \left\{ \lambda \mathbf{r}_{i}\right\} \right) .  \label{s22}
\end{eqnarray}%
If normalised $\Psi ^{\frac{^{\gamma }}{\lambda }}\left( \left\{ \mathbf{r}%
_{i}\right\} \right) $ yields $\rho \left( \mathbf{r}\right) $ then
normalised $\Psi ^{\frac{^{\gamma }}{\lambda }}\left( \left\{ \lambda
\mathbf{r}_{i}\right\} \right) $ yields $\lambda ^{3}\rho \left( \lambda
\mathbf{r}\right) .$ The Levy constrained minimization approach \cite%
{Levy:79} implies that normalised $\Psi ^{\frac{^{\gamma }}{\lambda }}\left(
\left\{ \lambda \mathbf{r}_{i}\right\} \right) $ yields $\lambda ^{3}\rho
\left( \lambda \mathbf{r}\right) $ and minimizes $\left\langle \Psi
\left\vert \hat{T}+\gamma \hat{V}_{ee}\right\vert \Psi \right\rangle _{\Psi
\rightarrow \lambda ^{3}\rho \left( \lambda \mathbf{r}\right) }.$ Therefore
\cite{ParrYang:bk89,DreizlerGross:bk90,JonesGunnarsson:89,LevyPerdew:85}$%
v^{\gamma }\left( \left[ \rho _{\lambda }\right] ;\mathbf{r}\right) =$ $%
\lambda ^{2}v^{\frac{^{\gamma }}{\lambda }}\left( \left[ \rho \right]
;\lambda \mathbf{r}\right) $ and%
\begin{equation}
\frac{\delta F^{^{\gamma }}\left[ \rho _{\lambda }\right] }{\delta \rho
_{\lambda }\left( \mathbf{r}\right) }+\lambda ^{2}v^{\frac{^{\gamma }}{%
\lambda }}\left( \left[ \rho \right] ;\lambda \mathbf{r}\right) =\mu
_{\lambda }^{\gamma },  \label{ip3}
\end{equation}%
and
\begin{eqnarray}
&&\left. \frac{d}{d\lambda }v^{\gamma }\left( \left[ \rho _{\lambda }\right]
;\mathbf{r}\right) \right\vert _{\lambda =1}  \notag \\
&=&2v^{\gamma }\left( \left[ \rho \right] ;\mathbf{r}\right) +\mathbf{%
r.\nabla }v^{\gamma }\left( \left[ \rho \right] ;\mathbf{r}\right) -\frac{%
\partial }{\partial \gamma }v^{\gamma }\left( \left[ \rho \right] ;\mathbf{r}%
\right)   \label{ip3aa}
\end{eqnarray}%
For an $N$-electron system,
\begin{equation}
\mu _{N,\lambda }^{\gamma }=E_{N}^{\gamma }\left( \lambda ^{2}v^{\frac{%
^{\gamma }}{\lambda }}\left[ \rho \right] \right) -E_{N-1}^{\gamma }\left(
\lambda ^{2}v^{\frac{^{\gamma }}{\lambda }}\left[ \rho \right] \right)
\label{ip3a}
\end{equation}%
with $E_{N}^{\gamma }\left( \lambda ^{2}v^{\frac{^{\gamma }}{\lambda }}\left[
\rho _{N}\right] \right) $ and $E_{N-1}^{\gamma }\left( \lambda ^{2}v^{\frac{%
^{\gamma }}{\lambda }}\left[ \rho _{N}\right] \right) $ the groundstate
energies of the $N$- and $\left( N-1\right) $- electron systems with the
same external potential $\lambda ^{2}v^{\frac{^{\gamma }}{\lambda }}\left[
\rho \right] .$ From Eq.(\ref{s22}) it follows that
\begin{eqnarray}
\mu _{N,\lambda }^{\gamma } &=&E_{N}^{\gamma }\left( \lambda ^{2}v^{\frac{%
^{\gamma }}{\lambda }}\left[ \rho \right] \right) -E_{N-1}^{\gamma }\left(
\lambda ^{2}v^{\frac{^{\gamma }}{\lambda }}\left[ \rho \right] \right)
\notag \\
&=&\lambda ^{2}\left( E_{N}^{\frac{\gamma }{\lambda }}\left( v^{\frac{%
^{\gamma }}{\lambda }}\left[ \rho \right] \right) -E_{N-1}^{\frac{\gamma }{%
\lambda }}\left( v^{\frac{^{\gamma }}{\lambda }}\left[ \rho \right] \right)
\right)   \notag \\
&=&\lambda ^{2}\mu   \label{ip3bb}
\end{eqnarray}%
where the last step follows from the definition of the chemical potential of
the $N$-electron system. Note that the chemical potential $\mu $ is
independent of $\gamma $\cite{LevyGorling:96,LevyGorlingb:96}$.$

The correlation energy $E_{c}^{\gamma }\left[ \rho \right] $ is defined as%
\cite{LevyPerdew:85,ParrYang:bk89,DreizlerGross:bk90}
\begin{eqnarray}
E_{c}^{\gamma }\left[ \rho \right] &=&\left\langle \Psi _{\rho }^{\gamma
}\left\vert \hat{T}+\gamma \hat{V}_{ee}\right\vert \Psi _{\rho }^{\gamma
}\right\rangle  \notag \\
&&-\left\langle \Psi _{\rho }^{0}\left\vert \hat{T}+\gamma \hat{V}%
_{ee}\right\vert \Psi _{\rho }^{0}\right\rangle ,  \label{ec1}
\end{eqnarray}%
where $\left\vert \Psi _{\rho }^{\gamma }\right\rangle $ is the ground state
wavefunction of $H^{\gamma }$ that yields $\rho $ and $\left\vert \Psi
_{\rho }^{0}\right\rangle $ the Kohn-Sham independent $N$-electron
groundstate wavefunction that yields the same density$.$ The correlation
part of the kinetic energy is given by%
\begin{equation}
T_{c}^{\gamma }\left[ \rho \right] =\left\langle \Psi _{\rho }^{\gamma
}\left\vert \hat{T}\right\vert \Psi _{\rho }^{\gamma }\right\rangle
-\left\langle \Psi _{\rho }^{0}\left\vert \hat{T}\right\vert \Psi _{\rho
}^{0}\right\rangle ,  \label{ec2}
\end{equation}%
and $\gamma V_{ee}^{\gamma }\left[ \rho \right] $\cite%
{LevyPerdew:85,ParrYang:bk89,DreizlerGross:bk90} can be written as%
\begin{equation}
\gamma V_{ee}^{\gamma }\left[ \rho \right] =\gamma E_{hx}\left[ \rho \right]
+E_{c}^{\gamma }\left[ \rho \right] -T_{c}^{\gamma }\left[ \rho \right] ,
\label{vee}
\end{equation}%
where $E_{hx}\left[ \rho \right] $ is the sum of the Hartree and exchange
energies. Note that $E_{hx}\left[ \rho \right] $ is independent of $\gamma .$
It follows from (\ref{ec1}) and (\ref{ec2}) that
\begin{equation}
t_{c}^{\gamma }\left( \mathbf{r}\right) +\gamma \frac{d}{d\gamma }%
v_{c}^{\gamma }\left( \mathbf{r}\right) -v_{c}^{\gamma }\left( \mathbf{r}%
\right) =0,  \label{dtc}
\end{equation}%
where $t_{c}^{\gamma }\left( \left[ \rho \right] ;\mathbf{r}\right) =\frac{%
\delta T_{c}^{\gamma }\left[ \rho \right] }{\delta \rho \left( \mathbf{r}%
\right) }$ and $v_{c}^{\gamma }\left( \left[ \rho \right] ;\mathbf{r}\right)
=\frac{\delta E_{c}^{\gamma }\left[ \rho \right] }{\delta \rho \left(
\mathbf{r}\right) }.$
Combining (\ref{e1}), (\ref{ip2}), (\ref{ip3}), (\ref{ip3aa}), (\ref{ip3bb})
and (\ref{dtc}) yields Eq. (\ref{ip4}).

\section{\label{appendixB}Derivation of Eq. (\ref{Vee})}

Multiply Eq.(\ref{ip4}) by the Fukui function and integrate over $\mathbf{r.%
}$ Use Eq.(\ref{bp7}), the fact that
\begin{equation}
\int d^{3}r^{\prime }\left( 3\rho \left( \mathbf{r}^{\prime }\right) +%
\mathbf{r.\nabla }\rho \left( \mathbf{r}^{\prime }\right) \right) =0
\label{b1}
\end{equation}%
and Eq.(\ref{dtc}) to arrive at
\begin{eqnarray}
&&\int d^{3}r^{\prime }\left[ 2v^{\gamma }\left( \left[ \rho \right] ;%
\mathbf{r}^{\prime }\right) +\mathbf{r.\nabla }v^{\gamma }\left( \left[ \rho %
\right] ;\mathbf{r}\right) \right] f^{\gamma }\left( \mathbf{r}\right) +
\notag \\
&&\int d^{3}r^{\prime }\left[ \gamma v_{hx}\left( \left[ \rho \right] ;%
\mathbf{r}^{\prime }\right) +v_{c}^{\gamma }\left( \left[ \rho \right] ;%
\mathbf{r}^{\prime }\right) -t_{c}^{\gamma }\left( \left[ \rho \right] ;%
\mathbf{r}^{\prime }\right) \right] f^{\gamma }\left( \mathbf{r}^{\prime
}\right)  \notag \\
&=&2\mu .  \label{b2}
\end{eqnarray}%
From the definition of $E_{c}^{\gamma }\left[ \rho \right] $ and $%
E_{c}^{\gamma }\left[ \rho \right] ,$ Eqs. (\ref{ec1}) and (\ref{ec2}) and
the fact that
\begin{equation}
E_{hx}\left[ \rho \right] =\left\langle \Psi _{\rho }^{0}\left\vert \hat{V}%
_{ee}\right\vert \Psi _{\rho }^{0}\right\rangle ,
\end{equation}%
it follows that
\begin{equation}
\gamma \frac{\delta V_{ee}^{\gamma }\left[ \rho \right] }{\delta \rho \left(
\mathbf{r}\right) }=\gamma v_{hx}\left( \left[ \rho \right] ;\mathbf{r}%
\right) +v_{c}^{\gamma }\left( \left[ \rho \right] ;\mathbf{r}\right)
-t_{c}^{\gamma }\left( \left[ \rho \right] ;\mathbf{r}\right) .  \label{b3}
\end{equation}%
Using the virial theorem \cite{LevyPerdew:85} and $\mu =E^{\gamma }\left[ \rho
_{N}\right] -E^{\gamma }\left[ \rho _{N-1}^{\gamma }\right] $,
\begin{eqnarray}
&&\int d^{3}r^{\prime }\left[ 2v^{\gamma }\left( \left[ \rho \right] ;%
\mathbf{r}^{\prime }\right) +\mathbf{r.\nabla }v^{\gamma }\left( \left[ \rho %
\right] ;\mathbf{r}\right) \right] f^{\gamma }\left( \mathbf{r}\right)
\notag \\
&=&2\mu -\gamma V_{ee}^{\gamma }\left[ \rho _{N}\right] +\gamma
V_{ee}^{\gamma }\left[ \rho _{N-1}^{\gamma }\right] .  \label{b4}
\end{eqnarray}%
Combining Eq. (\ref{b2}), (\ref{b3}) and (\ref{b4}) yields Eq. (\ref{Vee}).
For a detailed discussion see reference \cite{Joubert2011a}.

\bibliographystyle{apsrev}

\begin{thebibliography}{27}
\expandafter\ifx\csname natexlab\endcsname\relax\def\natexlab#1{#1}\fi
\expandafter\ifx\csname bibnamefont\endcsname\relax
  \def\bibnamefont#1{#1}\fi
\expandafter\ifx\csname bibfnamefont\endcsname\relax
  \def\bibfnamefont#1{#1}\fi
\expandafter\ifx\csname citenamefont\endcsname\relax
  \def\citenamefont#1{#1}\fi
\expandafter\ifx\csname url\endcsname\relax
  \def\url#1{\texttt{#1}}\fi
\expandafter\ifx\csname urlprefix\endcsname\relax\def\urlprefix{URL }\fi
\providecommand{\bibinfo}[2]{#2}
\providecommand{\eprint}[2][]{\url{#2}}

\bibitem[{\citenamefont{Hohenberg and Kohn}(1964)}]{HohenbergKohn:64}
\bibinfo{author}{\bibfnamefont{P.}~\bibnamefont{Hohenberg}} \bibnamefont{and}
  \bibinfo{author}{\bibfnamefont{W.}~\bibnamefont{Kohn}},
  \bibinfo{journal}{Phys. Rev. B} \textbf{\bibinfo{volume}{136}},
  \bibinfo{pages}{864} (\bibinfo{year}{1964}).

\bibitem[{\citenamefont{Ruzsinszky and Perdew}(2011)}]{AdriennRuzsinszky2011}
\bibinfo{author}{\bibfnamefont{A.}~\bibnamefont{Ruzsinszky}} \bibnamefont{and}
  \bibinfo{author}{\bibfnamefont{J.~P.} \bibnamefont{Perdew}},
  \bibinfo{journal}{Comput. Theor. Chem.} \textbf{\bibinfo{volume}{963}},
  \bibinfo{pages}{2} (\bibinfo{year}{2011}).

\bibitem[{\citenamefont{Perdew et~al.}(2010)\citenamefont{Perdew, Ruzsinszky,
  Csonka, Constantin, and Sun}}]{JohnP.Perdew2009}
\bibinfo{author}{\bibfnamefont{J.~P.} \bibnamefont{Perdew}},
  \bibinfo{author}{\bibfnamefont{A.}~\bibnamefont{Ruzsinszky}},
  \bibinfo{author}{\bibfnamefont{G.~I.} \bibnamefont{Csonka}},
  \bibinfo{author}{\bibfnamefont{L.~A.} \bibnamefont{Constantin}},
  \bibnamefont{and} \bibinfo{author}{\bibfnamefont{J.}~\bibnamefont{Sun}},
  \bibinfo{journal}{Phys. Rev. Lett.} \textbf{\bibinfo{volume}{103}},
  \bibinfo{pages}{026403} (\bibinfo{year}{2010}).

\bibitem[{\citenamefont{Csonka et~al.}(2009)\citenamefont{Csonka, Perdew,
  Ruzsinszky, Philipsen, Leb\`egue, Paier, Vydrov, and
  \'Angy\'an}}]{GaborI.Csonka2009a}
\bibinfo{author}{\bibfnamefont{G.~I.} \bibnamefont{Csonka}},
  \bibinfo{author}{\bibfnamefont{J.~P.} \bibnamefont{Perdew}},
  \bibinfo{author}{\bibfnamefont{A.}~\bibnamefont{Ruzsinszky}},
  \bibinfo{author}{\bibfnamefont{P.~H.~T.} \bibnamefont{Philipsen}},
  \bibinfo{author}{\bibfnamefont{S.}~\bibnamefont{Leb\`egue}},
  \bibinfo{author}{\bibfnamefont{J.}~\bibnamefont{Paier}},
  \bibinfo{author}{\bibfnamefont{O.~A.} \bibnamefont{Vydrov}},
  \bibnamefont{and} \bibinfo{author}{\bibfnamefont{J.~G.}
  \bibnamefont{\'Angy\'an}}, \bibinfo{journal}{Phys. Rev. B}
  \textbf{\bibinfo{volume}{79}}, \bibinfo{pages}{155107}
  (\bibinfo{year}{2009}).

\bibitem[{\citenamefont{Staroverov et~al.}(2004)\citenamefont{Staroverov,
  Scuseria, Tao, and Perdew}}]{Staroverov2004}
\bibinfo{author}{\bibfnamefont{V.~N.} \bibnamefont{Staroverov}},
  \bibinfo{author}{\bibfnamefont{G.~E.} \bibnamefont{Scuseria}},
  \bibinfo{author}{\bibfnamefont{J.}~\bibnamefont{Tao}}, \bibnamefont{and}
  \bibinfo{author}{\bibfnamefont{J.~P.} \bibnamefont{Perdew}},
  \bibinfo{journal}{Phys. Rev. B} \textbf{\bibinfo{volume}{69}},
  \bibinfo{pages}{075102} (\bibinfo{year}{2004}).

\bibitem[{\citenamefont{Tao et~al.}(2003)\citenamefont{Tao, Perdew, Staroverov,
  and Scuseria}}]{TaoPerdew03}
\bibinfo{author}{\bibfnamefont{J.}~\bibnamefont{Tao}},
  \bibinfo{author}{\bibfnamefont{J.~P.} \bibnamefont{Perdew}},
  \bibinfo{author}{\bibfnamefont{V.~N.} \bibnamefont{Staroverov}},
  \bibnamefont{and} \bibinfo{author}{\bibfnamefont{G.~E.}
  \bibnamefont{Scuseria}}, \bibinfo{journal}{Phys. Rev. Lett.}
  \textbf{\bibinfo{volume}{91}}, \bibinfo{pages}{146401}
  (\bibinfo{year}{2003}).

\bibitem[{\citenamefont{Perdew et~al.}(2005)\citenamefont{Perdew, Ruzsinszky,
  and Tao}}]{JohnP.Perdew2005}
\bibinfo{author}{\bibfnamefont{J.~P.} \bibnamefont{Perdew}},
  \bibinfo{author}{\bibfnamefont{A.}~\bibnamefont{Ruzsinszky}},
  \bibnamefont{and} \bibinfo{author}{\bibfnamefont{J.}~\bibnamefont{Tao}},
  \bibinfo{journal}{J. Chem. Phys.} \textbf{\bibinfo{volume}{123}},
  \bibinfo{pages}{062201} (\bibinfo{year}{2005}).

\bibitem[{\citenamefont{Harris and Jones}(1974)}]{HarrisJones:74}
\bibinfo{author}{\bibfnamefont{J.}~\bibnamefont{Harris}} \bibnamefont{and}
  \bibinfo{author}{\bibfnamefont{R.~O.} \bibnamefont{Jones}},
  \bibinfo{journal}{J. Phys. F} \textbf{\bibinfo{volume}{4}},
  \bibinfo{pages}{1174} (\bibinfo{year}{1974}).

\bibitem[{\citenamefont{Langreth and Perdew}(1975)}]{LangrethPerdew:75}
\bibinfo{author}{\bibfnamefont{D.~C.} \bibnamefont{Langreth}} \bibnamefont{and}
  \bibinfo{author}{\bibfnamefont{J.~P.} \bibnamefont{Perdew}},
  \bibinfo{journal}{Solid State Comm.} \textbf{\bibinfo{volume}{17}},
  \bibinfo{pages}{1425} (\bibinfo{year}{1975}).

\bibitem[{\citenamefont{Langreth and Perdew}(1977)}]{LangrethPerdew:77}
\bibinfo{author}{\bibfnamefont{D.~C.} \bibnamefont{Langreth}} \bibnamefont{and}
  \bibinfo{author}{\bibfnamefont{J.~P.} \bibnamefont{Perdew}},
  \bibinfo{journal}{Phys. Rev. B} \textbf{\bibinfo{volume}{15}},
  \bibinfo{pages}{2884} (\bibinfo{year}{1977}).

\bibitem[{\citenamefont{Gunnarson and Lundqvist}(1976)}]{GunnarsonLundqvist:76}
\bibinfo{author}{\bibfnamefont{O.}~\bibnamefont{Gunnarson}} \bibnamefont{and}
  \bibinfo{author}{\bibfnamefont{B.~I.} \bibnamefont{Lundqvist}},
  \bibinfo{journal}{Phys. Rev. B} \textbf{\bibinfo{volume}{13}},
  \bibinfo{pages}{4274} (\bibinfo{year}{1976}).

\bibitem[{\citenamefont{Levy}(1979)}]{Levy:79}
\bibinfo{author}{\bibfnamefont{M.}~\bibnamefont{Levy}}, \bibinfo{journal}{Natl.
  Acad. Sci. USA} \textbf{\bibinfo{volume}{76}}, \bibinfo{pages}{6062}
  (\bibinfo{year}{1979}).

\bibitem[{\citenamefont{Parr and Yang}(1989)}]{ParrYang:bk89}
\bibinfo{author}{\bibfnamefont{R.~G.} \bibnamefont{Parr}} \bibnamefont{and}
  \bibinfo{author}{\bibfnamefont{W.}~\bibnamefont{Yang}},
  \emph{\bibinfo{title}{Density Functional Theory of Atoms and Molecules}}
  (\bibinfo{publisher}{Oxford University Press}, \bibinfo{address}{New York},
  \bibinfo{year}{1989}).

\bibitem[{\citenamefont{Dreizler and Gross}(1990)}]{DreizlerGross:bk90}
\bibinfo{author}{\bibfnamefont{R.~M.} \bibnamefont{Dreizler}} \bibnamefont{and}
  \bibinfo{author}{\bibfnamefont{E.~K.~U.} \bibnamefont{Gross}},
  \emph{\bibinfo{title}{Density Functional Theory}}
  (\bibinfo{publisher}{Springer-Verlag}, \bibinfo{address}{Berlin},
  \bibinfo{year}{1990}).

\bibitem[{\citenamefont{Levy and Perdew}(1985)}]{LevyPerdew:85}
\bibinfo{author}{\bibfnamefont{M.}~\bibnamefont{Levy}} \bibnamefont{and}
  \bibinfo{author}{\bibfnamefont{J.~P.} \bibnamefont{Perdew}},
  \bibinfo{journal}{Phys. Rev. A} \textbf{\bibinfo{volume}{32}},
  \bibinfo{pages}{2010} (\bibinfo{year}{1985}).

\bibitem[{\citenamefont{G{\"o}rling and Levy}(1993)}]{GorlingLevy:93}
\bibinfo{author}{\bibfnamefont{A.}~\bibnamefont{G{\"o}rling}} \bibnamefont{and}
  \bibinfo{author}{\bibfnamefont{M.}~\bibnamefont{Levy}},
  \bibinfo{journal}{Phys. Rev. B} \textbf{\bibinfo{volume}{47}},
  \bibinfo{pages}{13105} (\bibinfo{year}{1993}).

\bibitem[{\citenamefont{Levy and
  G{\"o}rling}(1996{\natexlab{a}})}]{LevyGorling:96}
\bibinfo{author}{\bibfnamefont{M.}~\bibnamefont{Levy}} \bibnamefont{and}
  \bibinfo{author}{\bibfnamefont{A.}~\bibnamefont{G{\"o}rling}},
  \bibinfo{journal}{Phys. Rev. A} \textbf{\bibinfo{volume}{53}},
  \bibinfo{pages}{3140} (\bibinfo{year}{1996}{\natexlab{a}}).

\bibitem[{\citenamefont{Levy and
  G{\"o}rling}(1996{\natexlab{b}})}]{LevyGorlingb:96}
\bibinfo{author}{\bibfnamefont{M.}~\bibnamefont{Levy}} \bibnamefont{and}
  \bibinfo{author}{\bibfnamefont{A.}~\bibnamefont{G{\"o}rling}},
  \bibinfo{journal}{Phys. Rev. B} \textbf{\bibinfo{volume}{53}},
  \bibinfo{pages}{969} (\bibinfo{year}{1996}{\natexlab{b}}).

\bibitem[{\citenamefont{Berkowitz and Parr}(1988)}]{MaxBerkowitz1988}
\bibinfo{author}{\bibfnamefont{M.}~\bibnamefont{Berkowitz}} \bibnamefont{and}
  \bibinfo{author}{\bibfnamefont{R.~G.} \bibnamefont{Parr}},
  \bibinfo{journal}{J. Chem. Phys.} \textbf{\bibinfo{volume}{88}},
  \bibinfo{pages}{2554} (\bibinfo{year}{1988}).

\bibitem[{\citenamefont{P.~Fuentalba}(2007)}]{P.Fuentalba2007}
\bibinfo{author}{\bibfnamefont{C.~C.} \bibnamefont{P.~Fuentalba},
  \bibfnamefont{E.~Chamorro}}, \bibinfo{journal}{Int. J. Quantum Chem.}
  \textbf{\bibinfo{volume}{107}}, \bibinfo{pages}{37} (\bibinfo{year}{2007}).

\bibitem[{\citenamefont{Merzbacher}(1970)}]{Merzbacher1970}
\bibinfo{author}{\bibfnamefont{E.}~\bibnamefont{Merzbacher}},
  \emph{\bibinfo{title}{Quantum Mechanics}} (\bibinfo{publisher}{John Wiley},
  \bibinfo{year}{1970}).

\bibitem[{\citenamefont{Perdew et~al.}(1982)\citenamefont{Perdew, Parr, Levy,
  and Balduz}}]{PPLB:82}
\bibinfo{author}{\bibfnamefont{J.}~\bibnamefont{Perdew}},
  \bibinfo{author}{\bibfnamefont{R.}~\bibnamefont{Parr}},
  \bibinfo{author}{\bibfnamefont{M.}~\bibnamefont{Levy}}, \bibnamefont{and}
  \bibinfo{author}{\bibfnamefont{J.}~\bibnamefont{Balduz}},
  \bibinfo{journal}{Phys. Rev. Lett.} \textbf{\bibinfo{volume}{49}},
  \bibinfo{pages}{1691} (\bibinfo{year}{1982}).

\bibitem[{\citenamefont{Geerlings et~al.}(2003)\citenamefont{Geerlings, Proft,
  and Langenaeker}}]{P.Geerlings2003}
\bibinfo{author}{\bibfnamefont{P.}~\bibnamefont{Geerlings}},
  \bibinfo{author}{\bibfnamefont{F.~D.} \bibnamefont{Proft}}, \bibnamefont{and}
  \bibinfo{author}{\bibfnamefont{W.}~\bibnamefont{Langenaeker}},
  \bibinfo{journal}{Chem. Rev.} \textbf{\bibinfo{volume}{103}},
  \bibinfo{pages}{1793} (\bibinfo{year}{2003}).

\bibitem[{\citenamefont{Joubert}(2011)}]{Joubert2011a}
\bibinfo{author}{\bibfnamefont{D.~P.} \bibnamefont{Joubert}},
  \bibinfo{journal}{arXiv:1107.3219v1 [cond-mat.mtrl-sci], In press Phys. Rev.
  A}  (\bibinfo{year}{2011}).

\bibitem[{\citenamefont{Lieb}(1983)}]{Lieb1983}
\bibinfo{author}{\bibfnamefont{E.~H.} \bibnamefont{Lieb}},
  \bibinfo{journal}{Int. J. of Quantum Chem.} \textbf{\bibinfo{volume}{24}},
  \bibinfo{pages}{243} (\bibinfo{year}{1983}).

\bibitem[{\citenamefont{G{\"o}rling and Levy}(1995)}]{GorlingLevyB:95}
\bibinfo{author}{\bibfnamefont{A.}~\bibnamefont{G{\"o}rling}} \bibnamefont{and}
  \bibinfo{author}{\bibfnamefont{M.}~\bibnamefont{Levy}},
  \bibinfo{journal}{Int. J. Quantum Chem.} \textbf{\bibinfo{volume}{56}},
  \bibinfo{pages}{93} (\bibinfo{year}{1995}).

\bibitem[{\citenamefont{Jones and Gunnarsson}(1989)}]{JonesGunnarsson:89}
\bibinfo{author}{\bibfnamefont{R.}~\bibnamefont{Jones}} \bibnamefont{and}
  \bibinfo{author}{\bibfnamefont{O.}~\bibnamefont{Gunnarsson}},
  \bibinfo{journal}{Rew. Mod. Phys.} \textbf{\bibinfo{volume}{61}},
  \bibinfo{pages}{689} (\bibinfo{year}{1989}).

\end{thebibliography}

\end{document}